\documentclass[twoside]{article}
\usepackage{fleqn,espcrc2}

\usepackage{epsfig}

\def\eros{{\sc eros}}
\def\erod{{\sc eros2}}
\def\erou{{\sc eros1}}
\def\macho{{\sc macho}}
\def\ogle{{\sc ogle}}
\def\lmc{{\sc lmc}}
\def\smc{{\sc smc}}
\def\ie{{\em i.e.}}

\newcommand{\AmS}{{\protect\the\textfont2
  A\kern-.1667em\lower.5ex\hbox{M}\kern-.125emS}}

\title{Not enough stellar mass Machos in the Galactic halo}

\author{A.~Milsztajn
        and
        T.~Lasserre\address{~DSM, DAPNIA, 
        Service de Physique des Particules \\
        ~~~~CEA Saclay, F-91191 Gif-sur-Yvette Cedex, France} 
        \thanks{AMilsztajn@cea.fr, TLasserre@cea.fr},
        ~~on behalf of the EROS collaboration
        }
       
\begin{document}

%
%
%
%

\begin{abstract}

We present an update of results from the search for microlensing
towards the Large Magellanic Cloud (\lmc) by \eros\ 
(Exp\'erience de Recherche d'Objets Sombres).
We have now monitored 25 million stars over three years.  Because of the
small number of observed microlensing candidates (four), our results are 
best presented as upper limits on the amount of dark compact objects
in the halo of our Galaxy.  We discuss critically the candidates and 
the possible location of the lenses, halo or \lmc . We
compare our results to those of the {\sc macho} group.  Finally, we
combine these new results with those from our search towards the
Small Magellanic Cloud as well as earlier ones from the {\sc eros1}
phase of our survey.  The combined data is sensitive to compact
objects in the broad mass range $10^{-7} - 10 \, {\rm M}_{\odot}$.
The derived upper limit on the 
abundance of stellar mass {\sc macho}s rules out such objects 
as the dominant component of the Galactic
halo if their mass is smaller than  $2 {\rm M}_{\odot}$. 

\end{abstract}

\maketitle

\section{Research context}
        The search for gravitational microlensing in our Galaxy
has been going on for a decade, following the 
proposal to use this effect as a probe of the dark matter content
of the Galactic halo \cite{pac86}.  The first microlensing 
candidates were reported in 1993, towards the \lmc\ 
\cite{aub93,alc93} and the Galactic Centre \cite{uda93}
by the \eros , \macho\ and \ogle\ collaborations.

Because they observed no microlensing candidate with a duration
shorter than 10~days,
the \erou\ and \macho\ groups were able to exclude 
the possibility that more than 10\% 
of the Galactic dark matter resides in planet-sized objects
\cite{aub95,alc96,ren97,ren98,alc98}.  

However a few events were detected with longer time\-scales. 
In their two-year analysis \cite{alc97a}, the \macho\ group 
estimated, from 6-8 candidate events towards the \lmc , 
an optical depth of
order half that required to account for the dynamical mass of a 
``standard'' spherical dark halo\footnote{
$4 \times 10^{11}\:{\rm M}_\odot$ within 50~kpc};
the typical Einstein radius crossing time of the events, $t_E$, 
implied an average mass of about 0.5~M$_\odot$ for halo lenses 
\cite{alc97a}.
Based on two candidates, \erou\ set an upper limit on the 
halo mass fraction in objects of similar masses
\cite{ans96,ren97}, that is below that required to 
explain the rotation curve of our Galaxy\footnote{
Assuming the original two \erou\ candidates are microlensing events, 
they would correspond to an optical depth six times lower than that
expected from a halo fully comprised of \macho s.}.

The second phase of the \eros\ programme was started in 1996, with a 
ten-fold increase over \erou\ in
the number of monitored stars in the Magellanic Clouds.
The analysis of the first two years of data towards the 
Small Magellanic Cloud (\smc )
allowed the observation of one microlensing event \cite{pal98}
also detected by \cite{alc97b}.
This single event, out of 5.3 million monitored stars, allowed \erod\ 
to further constrain the halo composition, excluding in 
particular that more than 50~\% of the standard dark halo
is made up of $0.01 - 0.5 \:{\rm M}_\odot$ objects 
\cite{afo99}.  
In contrast, an optical detection of a halo
white dwarf population was reported \cite{iba99},
compatible with a galactic halo full of white dwarfs.

Very recently, the \macho\ group presented an analysis of 5.7 year
light curves 
of 10.7~million stars in the \lmc\ \cite{alc00} with an
improved determination of their detection efficiency and a better
rejection of background supernova explosions behind the \lmc .
They now favour a galactic halo \macho\ component of 20\% in the
form of 0.4~M$_\odot$ objects.  
Within a few days, the detection of a halo white dwarf population
at the level of a 10\% component was also reported by \cite{iba00}. 
Simultaneously, the \erod\ group
presented its results from a two-year survey of 17.5 million stars 
in the \lmc\ \cite{las00a}.  
One \erou\ microlensing candidate, {\sc eros1-lmc-2}, 
was seen to vary again, 8 years after its first brightening, and
was thus eliminated from the list of microlensing candidates.
Two new candidates were identified ({\sc eros2-lmc-3} and~4).  
Because this is much lower than expected if \macho s
are a substantial component of the galactic halo, and because these
two new candidates do not show excellent agreement with simple
microlensing light curves, \eros\ chose to combine these
results with those from previous \eros\ analyses, and
to quote an upper limit on the fraction of the galactic halo 
in the form of \macho s. 

In this article, we describe an update on the \erod\ \lmc\ data,
an analysis of the three-year light curves
from 25.5 million stars.  While the sensitivity is improved,
the main conclusions are unchanged compared to \cite{las00a}.
One of the two-year
candidates was seen to vary in the third season and was thus
rejected.  Three new candidates have been detected.   
We combine these \erod\ \lmc\ results with those of previous 
independent \eros\ analyses,  
and derive the strongest limit obtained thus far on the 
amount of stellar mass objects in the Galactic halo.

\section{Experimental setup and LMC observations}
	The telescope, camera, telescope operation and data reduction
are as described in \cite{bau97,pal98}. 
Since August 1996, we have been monitoring
66 one square-degree fields in the \lmc , 
simultaneously in two wide passbands. 
Of these, data prior to May 1999 from 39~square-degrees 
spread over 64~fields have been analysed.  In this period, two thirds
of the fields were imaged about 210 times in average; 
the remaining third were
imaged only about 110 times.  The exposure times range from 
3~min in the \lmc\ center to 12~min on the periphery; 
the average sampling is once every 4~days (resp. 8~days). 

\section{LMC data analysis}
        The analysis of the \lmc\ data set was done using a program
independent from that used in the \smc\ study, with largely
different selection criteria.
The aim is to cross-validate both programs 
(as was already done in the analysis
of \erou\ Schmidt photographic plates \cite{ans96})
and avoid losing rare microlensing events\footnote{
~We have checked that the present program finds the same \smc\
candidate as reported in \cite{pal98}.}.
The analysis is very similar to that reported in \cite{las00a}.
We only give here a list of the various steps, as well as a
short description of the differences with respect to our 
two-year analysis.   A detailed description of the analysis
will be provided in \cite{las00,las00b}.

We first select the 6\% ``most variable'' light curves, a sample
much larger than the number of detectable variable stars.
This subset of our data is ``enriched'' in genuine variable 
stars\footnote{~We monitor our selection efficiency with Monte-Carlo 
simulated variable star and microlensing light curves.}, 
but also and mainly
in photometrically biased light curves, {\it i.e.} those of
stars especially sensitive to the observing conditions, such as
stars very close to nebulosities or to bright stars.
Working from this ``enriched'' subset, we apply a first 
set of cuts to select, in each colour separately, 
the light curves that exhibit significant variations.
We first identify the baseline flux in the light curve - basically
the most probable flux.  
We then search for {\it runs} along the light curve,
\ie\ groups of consecutive measurements that are all on the same side 
of the baseline flux.
We select light curves that either
have an abnormally low
number of runs over the whole light curve, or
show one long run (at least 
5 valid measurements) that is very 
unlikely to be a statistical fluctuation. 
We then discard light curves with a low signal-to-noise ratio by 
requiring that the group of 5 most luminous
consecutive measurements be significantly further
from the baseline than the average spread of the measurements.
We also check that the measurements inside the most significant run
show a smooth time variation.

The second set of cuts compares the measurements with the best fit
point-lens point-source constant speed microlensing light curve
(hereafter ``simple microlensing'').  They
allow us to reject variable stars whose light curves differ too much
from simple microlensing, and are sufficiently loose not to reject 
light curves affected by blending, parallax 
or the finite size of the source, 
and most cases of multiple lenses or sources. 
We also require that the fitted time of maximum magnification
lie within the observing period or very close to it, and that
the fitted timescale is shorter than 300 days. The latter cut
is equivalent to requiring that the baseline flux of the star 
is observed for at least a few months; this is necessary in any
analysis using this baseline flux.  At this stage of the analysis,
all cuts have been applied independently in the two passbands.

\begin{figure} [ht] 
 \begin{center} \epsfig{file=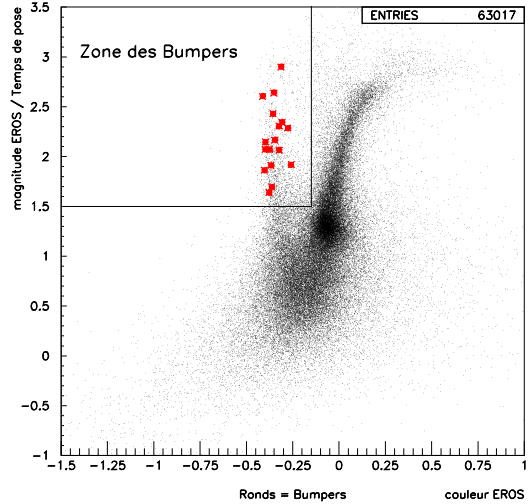,width=7.7cm}
  \caption{A colour-magnitude diagram in one \eros\ field, showing
   the location of candidates identified as ``blue bumpers'', either from
   their small amplitude, or from their chromatic variation
   (larger in the red than in the visible passband).
   }  
  \label{bumper}
 \end{center} 
\end{figure}

After this second set of cuts, stars selected separately 
in the two passbands represent
about 0.01\% of the initial sample; almost all of them
are found in two thinly populated zones of the colour-magnitude 
diagram. 
The third set of cuts deals with this physical background.
The first zone
contains stars brighter and much redder than those of the red clump;
variable stars in this zone are rejected if they vary by less than
a factor two or have a very poor fit to simple microlensing.  
The second zone is the top of the main sequence, where the
selected stars, known as blue bumpers \cite{alc97a}, 
display variations that are almost always smaller than 60\% of 
the base flux or at least 20\% lower in the visible passband 
than in the red one.  These cannot correspond to simple microlensing,
which is achromatic; neither can they correspond to microlensing
plus blending with another unmagnified star, 
as it would imply blending by even bluer stars,
which is very unlikely. 
We thus reject all candidates from the second zone
exhibiting one of these two features (see Fig.~\ref{bumper}).
%
%
%
%

Compared to the analysis in \cite{las00a}, two new cuts 
are introduced to reject other types of variable stars that were
not present in the two-year analysis.  The first one is aimed at
stars which have a roughly constant luminosity for some time, then
vary typically over one or two months to reach a new constant level.
We cannot yet conclude whether these are physical variable stars or
some kind of instrumental problem.  The second cut is aimed at 
nov\ae\ and supernov\ae . It rejects light curves which have a rise
time significantly smaller than the decline time; it is not applied to
events with a timescale longer than 60 days, in order not to 
reject microlensing phenomena with parallax
effects, that also show an asymmetry.

The final cuts are simply tighter cuts on the fit quality, applied
to both colours (whereas similar previous cuts were applied independently
in each passband), and a requirement that the observed magnification
be at least 1.40~.

The tuning of each cut and the calculation of the microlensing 
detection efficiency are done with 
simulated simple microlensing light curves, as described in 
\cite{pal98}. 
For the efficiency calculation, 
microlensing parameters are drawn as follows~: 
time of maximum magnification $t_0$ uniformly 
within the observing period $\pm 150$~days,
impact parameter normalised to the Einstein
radius $u_0 \in [0,2]$ uniformly, and timescale 
$t_E \in [1,400]$ days uniformly in $\ln(t_E)$.  
All cuts on the data were also applied to the simulated
light curves.

%
%
\begin{figure} [ht] 
 \begin{center} \epsfig{file=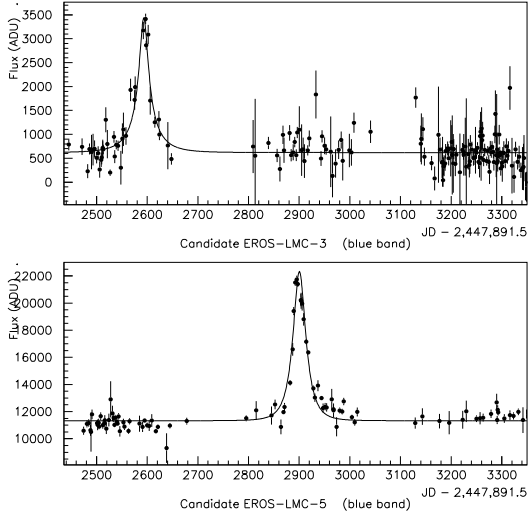,width=7.7cm}
  \caption{Light curves of candidates EROS-LMC-3 and 5 (visible passband). 
   The plain curves show the best point-lens point-source fits;
   time is in days since Jan 1.0, 1990 (JD 2,447,892.5).}  
  \label{cdl1_evts}
 \end{center} 
\end{figure}
\begin{figure} [ht] 
 \begin{center} \epsfig{file=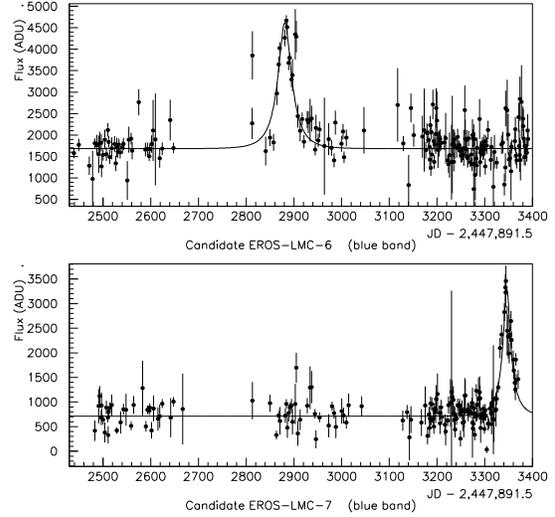,width=7.7cm} 
  \caption{Same as Fig.2 for candidates EROS-LMC-6 and 7. 
   }  
  \label{cdl2_evts}
 \end{center} 
\end{figure}
%
%

Only four candidates remain after all cuts. 
Of the two candidates presented in \cite{las00a},
{\sc eros2-lmc-3} is still a member of this list, 
while {\sc eros2-lmc-4} was seen to vary at least twice in the third
season and was thus rejected.  There are three new candidates,
numbered 5 to 7.
Their light curves are shown
in Figs.~\ref{cdl1_evts} and \ref{cdl2_evts}; microlensing fit parameters
are given in Table~\ref{eventparm}. 
Although the candidates pass all cuts, agreement with simple 
microlensing is not excellent.  In particular, 
{\sc eros2-lmc-5} is dubious~:
it has a bad fit to simple microlensing and is located
in an atypical region of the colour-magnitude diagram.
The geometric mean of the candidates timescales is about 32~days,
including that of the \erou\ candidate {\sc lmc-1}.

%
%
\begin{table}[ht]
\begin{center} \vspace{-0.0cm} 
\begin{tabular}{l|cccccc}
\hline
   &$u_0$&$t_E$&$c_{\rm \,bl}^R$&$c_{\rm \,bl}^V$&$\chi^2/{\rm dof}$& 
$V_J$ \\
\hline
 \lmc-3 & $0.21$&$44$&$0.75$& 1 &219/143&22.4\\
\hline
 \lmc-5 & $0.58$&$24$&$0.91$& 1 &658/176&19.2\\
\hline
 \lmc-6 & $0.38$&$36$&$0.72$& 1 &682/411&21.3\\
\hline
 \lmc-7 & $0.23$&$33$&$0.45$& 1 &722/356&22.7\\
\hline

\end{tabular}
%
%
%
\caption{Results of microlensing fits to the four \erod\ \lmc\ candidates;
 $t_E$ is the Einstein radius crossing time in days, 
$u_{0}$ the impact parameter, and $c_{\rm \,bl}^{R(V)}$ the 
$R(V)$ blending coefficients, constrained to be smaller than unity.
}
\label{eventparm}
\end{center} \vspace{-0.5cm}
\end{table}
%
%

The microlensing detection efficiency of this analysis, 
normalised to events with an impact 
parameter $u_0<1$ and to an observing period $T_{\rm obs}$ of three
years, is summarised in Table~\ref{eff}.
The main source of systematic error is the uncertainty in the 
influence of blending. Blending lowers the observed magnifications
and timescales.  While this decreases the efficiency for a given star, 
the effective number of monitored stars is increased so that there
is partial compensation.
This effect was studied with synthetic images using measured magnitude
distributions \cite{pal97}. 
Our final efficiency is within 10\% of the ``naive'' sampling efficiency.
Compared to the efficiency in \cite{las00a}, the present
one is improved for the longest and shortest  durations, 
but slightly lower for average durations around 50~days.  
This is largely explained by the fact that 
we have included in the present analysis stars in external \lmc\
fields that were sampled less frequently.
%
%
\begin{table}[ht]
\begin{center} \vspace{-0.0cm} 
\vspace{-0.0cm}
 \begin{tabular}{c|cccccccc} \hline
     $t_E$ & 6.3& 13 & 28& 40& 90& 175& 250 & 360\\ \hline
$\epsilon$ & 2.7& 6.7& 11& 14& 19&  22& 17.5&  2 \\ \hline
\end{tabular}
%
%
\caption{\erod\ detection efficiency in \% for the \lmc\ 3-year
analysis, as a function of the
Einstein radius crossing time $t_E$ in days, 
normalised to events generated with $u_0<1$, and
to $T_{\rm obs}=3{\rm \: yrs}$. 
}
\label{eff} 
\end{center} \vspace{-0.6cm}
\end{table}
%
%

\section{Limits on Galactic halo MACHOs}
    \eros\ has observed six microlensing candidates towards the
Magellanic Clouds, one from \erou\ and four from \erod\ towards
the \lmc , and one towards the \smc .
As discussed in \cite{pal98},
and further in \cite{gra00},
we consider that the long duration
of the \smc\ candidate together with the absence of any detectable
parallax, in our data as well as in that of the \macho\ group
\cite{alc97b}, indicates that it is most likely
due to a lens in the \smc \footnote{
~Alternatively, it can be argued that, if due to a galactic halo 
lens, this event corresponds to a heavy lens \cite{alc97b,afo99,gra00}.
}.  For that reason, the limit
derived below uses the five \lmc\ candidates.  (The limit 
corresponding to all six candidates would be about a factor 1.13
times the limit shown, for masses larger than 0.01 solar mass.)
 
The limits on the contribution of dark compact objects
to the Galactic halo are obtained by comparing the number 
and durations of 
microlensing candidates with those expected from Galactic
halo models.   
We use here the so-called ``standard'' halo model described in
\cite{pal98} as model 1, but have checked
that we obtain similar results for other reasonable halo models.
The model predictions are computed for each \eros\
data set in turn, taking into account the corresponding detection 
efficiencies
(\cite{ans96,ren98,afo99} and Table~\ref{eff} above),
and the four predictions are then summed.
In this model, all dark objects have the same mass $M$; we have
computed the model predictions for many trial masses $M$ in turn,
in the range [$10^{-8}\:{\rm M}_\odot$, $10^2\:{\rm M}_{\odot}$].

The method used to compute the limit is as  
in \cite{ans96}. We consider two ranges of
timescale $t_E$, within or outside the interval 
I~$ = [ 7.5 ; 190 ]$~days.  This interval was chosen as follows.
We first determine the average mass corresponding to the mean duration
of the five \lmc\ candidates, at about $0.2 \:{\rm M}_{\odot}$.
We then compute the expected distribution of microlensing
timescales for this average mass and check that the observed spread 
in timescales for the candidates is compatible with the width of
this distribution.  This means that our candidates are compatible 
with the hypothesis that their spread in mass contributes very little
to the width of the timescale distribution.
The interval~I is then chosen as a symmetrical interval in $\ln(t_E)$
that contains 99\% of the timescale distribution for halo
\macho s of $0.2 \:{\rm M}_{\odot}$.  Of course, all five \lmc\ 
candidates have timescales well within the interval I. 

We can then compute, for each mass
$M$ and any halo fraction $f$,
the combined Poisson probability for obtaining,
in the four different \eros\ data sets taken as a whole, 
zero candidate outside I and 
five or less 
within I.  For any value of $M$,
the limit $f_{\rm max}$ is the value of $f$ for which this probability
is 5\%.  Whereas the actual limit depends somewhat on the precise
choice of I, the difference (smaller than 5\%) is noticeable 
only for masses around 0.01 and $10 \:{\rm M}_{\odot}$. 
Furthermore, we consider our choice for I to be a conservative one.
%
%
\begin{figure} [ht] 
 \begin{center} \epsfig{file=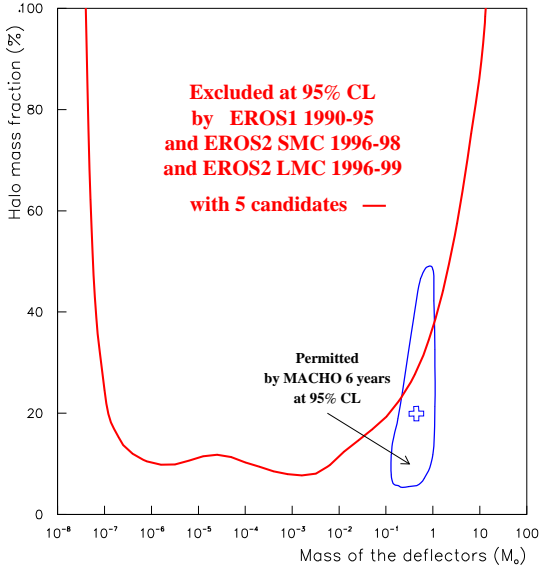,width=7.7cm}
  \vspace{-.1cm} \caption{ 95\% C.L. exclusion diagram on the halo mass
  fraction in the form of compact objects of mass $M$, for the
  standard halo model ($4 \times 10^{11}\:{\rm M}_\odot$ inside
  50~kpc), from all \lmc\ and \smc\ \eros\ data 1990-99.  
  The solid line is the limit inferred from the five \lmc\ microlensing 
  candidates.
  The new {\sc macho} 95\%
  C.L. accepted region is the closed contour, with the preferred 
  value indicated by the cross \cite{alc00}.} 
  \label{excl}
  \end{center} \vspace{-0.5cm}
\end{figure}
%
%

Figure \ref{excl} shows the 95\% C.L. exclusion limit derived 
from this analysis on the halo mass fraction, $f$,
for any given dark object mass, $M$. 
The solid line corresponds to the five \lmc\ candidates;
it is the main result of this article. 
This limit rules out a standard spherical halo model fully comprised of 
objects with any mass function inside the range 
$[10^{-7}-10] \; M_{\odot}$.
In the region of stellar mass objects, where this result
improves most on previous ones, the new \lmc\ data contribute
about 73\% to our total sensitivity (the \smc\ and \erou\ 
\lmc\ data contribute 10\% and 17\% respectively).  
The total sensitivity for $t_E = 50$~days, 
that is proportional to the sum of 
$N_* \, T_{\rm obs} \, \epsilon (t_E = 50{\rm d})$
over the four \eros\ data sets,
is about 3.2 times larger than that of \cite{alc97a}
and two thirds that of \cite{alc00}.

\section{Discussion}
        After nine years of monitoring the Magellanic Clouds, 
\eros\ has a meager crop of five microlensing candidates towards 
the \lmc\ and one towards the \smc , whereas about 30 events are expected
towards the \lmc\
for a spherical halo fully comprised of $0.4 \:{\rm M}_\odot$ objects.
Moreover, some of the candidates cannot be considered excellent.
These candidates were obtained from four different data 
sets analysed by independent, cross-validated programs.  
So, the small number of observed events is unlikely
to be due to bad (and overestimated) detection efficiencies.

This allows us to put strong constraints on the fraction
of the halo made of objects in the range [$10^{-7}\:{\rm M}_\odot$,
$10\:{\rm M}_{\odot}$], excluding in particular at the 95\% C.L. that 
more than 40\% of the standard halo be made of objects with up to
$1 \:{\rm M}_\odot$.  
The preferred value quoted in \cite{alc00},
$f = 0.2$ and $0.4\:{\rm M}_\odot$,
is consistent with our limit as can be seen in Fig.~\ref{excl}.
(The upper part -~about 25\%~- of the domain allowed by 
\cite{alc00} is excluded by the limit we report here.)

There are several differences which
should be kept in mind while comparing the two experiments.
First, \eros\ uses less crowded fields than \macho\ with the result
that blending is relatively unimportant for \eros .
(Were \eros\ results to be corrected for blending, the detection 
efficiency would increase slightly and the reported limit
would be stronger.)
Second, \eros\ covers a larger solid angle (64~deg$^2$ in the \lmc\
and 10~deg$^2$ in the \smc ) than \macho , which monitors primarily
15~deg$^2$  in the central part of the \lmc .
The \eros\ rate should thus be less contaminated by self-lensing,
i.e. microlensing of \lmc\ stars by dimmer \lmc\ objects,
which should be more common in the central regions.
The importance of self-lensing was first stressed 
in \cite{wu94,sa94}.
Third, the \macho\ data have a more frequent time sampling.

The results from \eros\ and \macho\ are apparently consistent,
but the way they are interpreted is different.  \macho\ reports 
a signal and considers the contamination of its sample 
as low or null.  \erod\ quotes an upper limit and does not claim
its sample to be background-free.  The position
of the lenses along the line of sight, halo or Magellanic Clouds, 
is also an issue.  \macho\ has compared the spatial
distribution of its candidates across the face of the \lmc\
and observes a better agreement with the halo hypothesis than
with a specific model of the \lmc .  On the other hand,
because the \eros\ stars are spread over a wider field,
the fact that the \eros\ sample corresponds to a lower central
value of the event rate (about twice lower than that of \macho )
is compatible with an interpretation where a noteable fraction
of the events are due to self-lensing.  The small number of
\eros\ candidates precludes at present any definitive conclusion
on that topic.

It seems likely that the single most important input to the question
of the position of the lenses will come from the comparison
of the microlens candidates samples towards the \smc\ 
with those towards the \lmc .  Because the two lines of sight are
rather close (about 20~degrees apart), 
the timescale distributions of microlensing candidates towards
the two Clouds should be nearly identical if lenses belong to
the galactic halo.  Also, the event rates should be comparable, 
although the ratio is more halo model dependent. 
At present, \eros\ has analysed
two seasons of \smc\ data \cite{afo99} and \macho\
has not yet presented its detection efficiency towards the
\smc .  From the published \eros\ efficiencies, and assuming
that the \macho\ efficiencies towards the \smc\ are similar 
to those towards the \lmc , it can be expected that the 
completed experiments will have gathered between five and ten
microlenses towards the \smc .  This should allow a significant
comparison of the timescales (see also the discussion in \cite{gra00}).

Finally, let us mention that,
given the scarcity of our candidates 
and the possibility that some observed microlenses
actually lie in the Magellanic Clouds, 
\eros\ is not willing at present to quote a non zero 
{\it lower} limit on the fraction of the Galactic halo comprised of
dark compact objects with masses up to a few solar masses.

\bigskip
{\bf Acknowledgements}
We are grateful to D. Lacroix and the staff at the Observatoire de
Haute Provence and to A. Baranne for their help with the MARLY
telescope.  
The support by the technical staff at ESO, La Silla, 
is essential to our project.
We thank J.F. Lecointe for assistance with the online computing.
We thank all non-\eros\ members who participated in data taking.

\end{document}